\documentclass[dvips,twoside,fleqn]{article}
\usepackage{times}
\usepackage{epsfig}
\usepackage{amstext}
\usepackage{amssymb}
\usepackage{espcrc2}

\title{The $(\lambda \Phi^4)_4$ theory on the lattice:
effective potential and triviality.}
\author{
A. Agodi\address{Dept. of Physics - Univ. of Catania - Corso Italia 57 - 95129 Catania - Italy}, 
G. Andronico\address{INFN - Sezione di Catania - Corso Italia 57 - 95129 Catania - Italy},
P. Cea\address{INFN - Sezione di Bari - Via Amendola 173 - 70126 Bari - Italy},
M. Consoli$^{\text{b}}$
and L. Cosmai$^{\text{c}}$
}

\begin{document}

\begin{abstract}
We compute numerically the effective potential for the $(\lambda \Phi^4)_4$ theory on the lattice. Three
different methods were used to determine the critical bare mass for the chosen bare coupling value. Two
different methods for obtaining the effective potential were used as a control on the results. We compare our
numerical results with three theoretical descriptions. Our lattice data are in quite good agreement with the
``Triviality and Spontaneous Symmetry Breaking'' picture.

\end{abstract}
\maketitle

\section{Introduction.}

The response to an external field is a standard tool to obtain 
non-perturbative information in 
quantum field theories. This is even more important in a scalar field theory
since the 
introduction of an external current gives directly the effective potential 
(up to an integration constant), thus providing important information on the stability of the
system.
 
The conventional interpretation of the triviality in $(\lambda\Phi^4)_4$ theories
is based on Renormalization-Group-Improved-Perturbation-Theory (RGIPT)~\cite{RGIPT1,RGIPT2,RGIPT3}. 
This picture predicts
a second-order phase transition and
a vanishing Higgs mass $m_h$ 
in the continuum limit if $v$, the physical v.e.v, is held fixed.
An alternative interpretation of triviality~\cite{C&S,Agodi1,Agodi2} predicts
a phase transition that is very weakly 
first-order and that 
$m_h$ and $v$ are both finite, cutoff-independent 
quantities.
The latter interpretation originates~\cite{C&S}  
from the Gaussian approximation.
The two alternative pictures can be distinguished by a sufficiently precise lattice 
calculation of the effective potential $V_{\text{eff}}$ to this end
we perform a model-independent ``numerical experiment'' to test the predictions of both 
the conventional RGIPT picture and the alternative picture.

\section{Effective potential on the lattice.}

To evaluate the effective potential on the lattice
we introduce a constant external source $J$ in 
the classical action for the  (one-component) $(\lambda\Phi^4)_4$ 
theory in Euclidean space.
One of the possible lattice discretizations reads 
\begin{eqnarray}
\label{lattactionJ}
\lefteqn{
S =\sum_x \left\{ \frac{1}{2} \sum_\mu \left[ \Phi(x+\hat e_{\mu}) - 
\Phi(x) \right]^2 + \right. 
}  \nonumber  \\ 
\lefteqn{
\left. \qquad  \qquad \frac{r_0}{2} \Phi^2(x)  +
\frac{\lambda_0}{4} \Phi^4(x) -  J \Phi(x)  \right\}
}
\end{eqnarray}
where $x$ stands for a generic lattice site, the lattice fields are expressed in 
lattice units ($a=1$), and $\lambda_0 >0$.
For SSB the basic quantity is the expectation 
value of the bare scalar field $\Phi(x)$ (B=Bare)
$\langle \Phi \rangle _J = \phi_B(J)$,   
since determining $\phi_B(J)$ at several $J$-values is equivalent 
to inverting the relation
$J=J(\phi_B)= dV_{\text{eff}}/d\phi_B$   
%
%
involving the effective potential $V_{\text{ eff}}(\phi_B)$.
In this way, 
starting from the action in Eq.(\ref{lattactionJ}), the effective potential of 
the theory is {\it rigorously} defined up to an arbitrary integration 
constant (usually chosen to fix $V_{\text{eff}} (0)=0$) and is convex downward~\cite{effpot}.
  
Since we want to simulate the $(\lambda \Phi^4)_4$ lattice theory
close to the continuum limit
we have to determine the critical value $r_c$
of the bare mass parameter $r_0$ at $J=0$, which
defines the `Coleman-Weinberg regime'
where $m$, the mass gap of the symmetric phases, vanishes.
The critical bare mass parameter $r_c$ can be determined from the 
susceptibility $\chi= 1/\Omega_{\mathrm{latt}} \left[ \left\langle \Phi^2 \right\rangle - 
\left\langle \Phi \right\rangle^2 \right]  $
%
%
where  $\Phi = 1/\Omega_{\mathrm{latt}} \sum_x  \Phi(x)$ 
%
%
is the average field for a given configuration and the brackets stand for the average 
on the lattice configurations produced in a Monte Carlo run.  

One expects that, near the critical region, 
$\chi^{-1} \sim (r_c-r_0)$, modulo logarithmic corrections to the 
free-field scaling law.  One can thus determine $r_c$ by extrapolation 
to vanishing $\chi^{-1}$.  
Strictly speaking, this method is valid only 
for a second-order phase transition where the phase transition value $r_0 \equiv r_s$ 
coincides with $r_c$ and such that, at 
$r_c=r_s$, both $m$ and $m_h$ vanish.  In the case of a very weak first-order 
phase transition where~\cite{C&S} 
$|r_c-r_s|/r_c \sim \exp(-1/(2s))$, $s \equiv 3\lambda_0/16\pi^2 \ll 1$, the induced 
numerical uncertainty should be negligible.

For the Monte Carlo simulation 
of the lattice field theory described by Eq.~(\ref{lattactionJ})
we followed the upgrade of the scalar field $\Phi(x)$ (using Metropolis) by the upgrade of 
the sign of $\Phi(x)$ according to the embedded Ising dynamics~\cite{embedded}.
The zero external field spin-flip probability of the Swendsen-Wang algorithm is slightly modified 
to take into account the external current  $J$.

Our data~\cite{Agodi2} are well described by the 
simple linear fit  $\chi^{-1}=a |r-r_c|$
%
%
and do not show evidence of the logarithmic corrections.
We evaluate the above quantity both in the symmetric and broken phase. 
We determined $r_c$ also through the following fit~\cite{Hasenfratz}
to the generalized magnetization $\langle \Phi \rangle$:
\begin{equation}
\label{hasenfit}
\langle \Phi \rangle = \alpha (r_c - r)^{1/2} | \ln|r-r_c| |^\beta + \delta 
\,.
\end{equation}
Combining the above 
estimates for  $r_c$ we get  $r_c=-0.2280 (9)$, in perfect
agreement with the independent analysis~\cite{brahm}
which predicts, for $\lambda_0=0.5$ and
$L=16$, $r_c=-0.2279(10)$. 
Thus we have {\it three} 
independent and consistent evaluations of 
$r_c$ at $\lambda_0=0.5$ on a $16^4$ lattice, that
represent a precise 
input definition of the 
`Coleman-Weinberg regime' with 
the action~Eq.(\ref{lattactionJ}) at $J=0$.
%
%
%
%
%
%
%
%
%
%
%
%
%
%



We have used two independent methods to compute the 
effective potential.  
Firstly, we ran simulations of the lattice action Eq.~(\ref{lattactionJ}) for 
16 different values of the external source in the range 
$0.01\leq |J| \leq 0.70$.  In this way, as outlined in Eqs.~(2-4), 
we directly obtain the slope of the effective potential (from which 
$V_{\mathrm{eff}}$ can be obtained, up to an additive integration constant).
We performed our numerical simulations
by using for the sign upgrade both the 
Metropolis and the S-W cluster algorithm.
\begin{table*}
\tabcolsep .2cm
\renewcommand{\arraystretch}{1.2}
\begin{center}
\begin{tabular}{|c||l|c|l|}
\hline
 data    & $J^{\text{triv}}$  & $ J^{\text{2-loop}}$
     & $J^{\text{ID}}$ \\ \hline \hline
Metropolis & $\alpha=0.0152 (2)$  & $\mu=8.0304 (449)$
& $|\mu|=2.70 (1) \times 10^8$        \\
         & $\gamma=0.4496 (1)$    & $M^2=-0.0025 (1)    $    
& $A=-0.0055 (3)$  \\
         & $\chi^2=\frac{15}{16-2}$ & $\chi^2=\frac{142}{16-2}$ 
& $\chi^2=\frac{116}{16-2}$
\\ \hline \hline
Swendsen-Wang  & $\alpha=0.0152 (1)$      & $\mu=8.0128 (283)$
& $|\mu|=2.70 (1) \times 10^8$        \\
         & $\gamma=0.44962 (4)$   & $M^2=-0.00249 (7)    $      
& $A=-0.0055 (2)$  \\
         & $\chi^2=\frac{13}{16-2}$  & $\chi^2=\frac{284}{16-2}$  
& $\chi^2=\frac{223}{16-2}$
\\ \hline \hline
constraint eff. pot. & $\alpha=0.0156 (2)$ & $\mu=7.9883 (455)$
& $|\mu|=2.69 (1) \times 10^8$        \\
         & $\gamma=0.4494 (1)$      & $M^2=-0.0028 (1)$           
& $A=-0.0063 (3)$  \\
         & $\chi^2=\frac{10}{16-2}$ & $\chi^2=\frac{109}{16-2}$ 
& $\chi^2=\frac{85}{16-2}$ 
\\ \hline
\end{tabular}
\caption{Results of the fits of  $J^{\text{triv}}$, $J^{\text{2-loop}}$, and $J^{\text{ID}}$ to our 
three different sets of  lattice data.}
\end{center}
\label{table:I}
\end{table*}

As a third additional check of our results, we performed a calculation 
using an alternative approach~\cite{constraint} 
to $V_{\mathrm{eff}}$  based on the approximate effective potential 
$U_{\rm eff} (\phi_B;\Omega)$ defined through
\begin{eqnarray}
\label{eq:3.1}
\lefteqn{
\exp\left\{-U_{\mathrm{eff}}(\phi_B;\Omega) \right\} = 
} \nonumber \\  
\lefteqn{
\int [D\Phi]
~\delta \left( {{1}\over{\Omega}}\int \!d^4x~ \Phi(x)-\phi_B \right) 
\exp{-S[\Phi]}
\, .
}
\end{eqnarray}
In the limit in which the 4-volume $\Omega \to \infty$, 
$U_{\rm eff}$
tends to the exact 
$V_{\rm eff}(\phi_B)$.  
The difference between $U_{\rm eff}(\phi_B;\Omega)$ and
$V_{\rm eff}(\phi_B)$
gives both a consistency check of our calculations and
an indication of the effects due to the finiteness of our lattice.

\section{Comparing theory with the lattice data.}

We can now compare our three different sets of lattice data with the 
existing theoretical expectations.  

In the case of the  picture in which
triviality is compatible with SSB~\cite{C&S,Agodi2}
the predicted form (in the Coleman-Weinberg case, $r_0=r_c$, where no 
quadratic term is present in the effective potential) is:
\begin{equation}
\label{triviality}
J^{\rm triv}(\phi_B)={{dV_{\rm triv}}\over{d\phi_B}}=
\alpha \phi^3_B \ln (\phi^2_B) + \gamma \phi^3_B, 
\end{equation}
where $\alpha$ and $\gamma$ are free parameters.  (Their values are
approximation-dependent within the class of ``triviality-compatible'' 
approximations.)  

The RGIPT prediction exists in various slightly different forms in the 
literature.  We have first used the full two-loop calculation of Ford and 
Jones~\cite{Jones}  in the dimensional regularization scheme.  
The theoretical prediction for $J^{\rm 2-loop}(\phi_B)=dV^{\rm 2-loop}/d\phi_B$
depends on two free parameters:
the 't~Hooft scale $\mu$ and the mass 
parameter $M^2$ of the classical potential.

A different version of the RGIPT prediction, which re-sums various 
terms, is given in Eq.~(242), Sect.~5.4.2, of the textbook~\cite{drouffe} by Itzykson 
and Drouffe (ID), namely 
\begin{equation}
\label{itzdrouffe}
J^{\rm ID}(\phi_B)=\frac{A\phi_B}{ 
\left| \ln \frac{|\mu|}{|\phi_B|} \right|^{1/3} } +
{{(4\pi)^2 \phi^3_B}\over{18  \ln{{|\mu|}\over{|\phi_B|}}  }} 
\,,
\end{equation}
and again we have two free parameters $A$ and $\mu$.  
 
The results of fitting the lattice data to the three theoretical predictions,
are shown in Table 1.


\section{Conclusions.}
We have performed  a numerical experiment to test  two 
basically different and alternative pictures of `triviality'
in the `Coleman-Weinberg-regime'.
Our results are striking and confirm that the lattice data cannot be reproduced 
by all theoretical models. Indeed
the results of fitting the 
conventional theoretical predictions to our lattice data 
gives 
totally unacceptable values of the $\chi^2/f$.
On the other hand Eq.~(\ref{triviality}), based on the 
alternative picture, 
gives 
excellent fits to all sets of data. 
This is an 
important evidence
for the 
unconventional interpretation of  ``triviality''
of Refs.~\cite{C&S,Agodi1,Agodi2}.


\end{document}